\documentclass[conference]{IEEEtran}
\IEEEoverridecommandlockouts

\usepackage{cite}
\usepackage{amsmath,amssymb,amsfonts}
\usepackage{algorithmic}
\usepackage{graphicx}
\usepackage{textcomp}
\usepackage{xcolor}
\usepackage{subfigure}
\def\BibTeX{{\rm B\kern-.05em{\sc i\kern-.025em b}\kern-.08em
    T\kern-.1667em\lower.7ex\hbox{E}\kern-.125emX}}
\begin{document}

\title{Illinois Express Quantum Network for Distributing and 
Controlling Entanglement on Metro-Scale}

\author{
Wenji Wu$^{1}$, 
Joaquin Chung$^{2}$, 
Gregory Kanter$^{3}$, 
Nikolai Lauk$^{4}$, 
Raju Valivarthi$^{4}$, \\
Russell R. Ceballos$^{1,5}$,
Cristián Peña$^{1,4}$, 
Neil Sinclair$^{4,6}$, 
Jordan M. Thomas$^{7}$, 
Ely M. Eastman$^{7}$, \\
Si Xie$^{4}$,
Rajkumar Kettimuthu$^{2}$, 
Prem Kumar$^{7}$, 
Panagiotis Spentzouris$^{1}$, 
Maria Spiropulu$^{4}$
\\
$^{1}$Fermi National Accelerator Laboratory \\
$^{2}$Argonne National Laboratory \\
$^{3}$NuCrypt LLC\\ 
$^{4}$California Institute of Technology \\
$^{5}$Chicago State University \\
$^{6}$Harvard University \\
$^{7}$Northwestern University

}

\maketitle

\begin{abstract}

We describe an implementation of a quantum network over installed fiber 
in the Chicago area. We present network topology and control 
architecture of this network and illustrate preliminary results for 
quantum teleportation and coexistence of quantum and classical data on 
the same fiber link.

\end{abstract}


\section{Introduction} \label{sec:intro}
Analogous to classical optical networks, eventually quantum optical networks will be used to interconnect quantum processors including quantum computers and sensors. While it is attractive to leverage as much of the traditional fiber-optic infrastructure as possible, quantum signals have different characteristics that, for instance, preclude their use with optical amplifiers or electrical termination before the end-point. Additionally, they are very sensitive to loss and even miniscule levels of added noise. As such, we expect quantum networks to require custom engineering while also benefiting from as much compatibility with traditional networks as is practical.

The Illinois Express Quantum Network (IEQNET) is a program for developing metro-scale quantum networking over deployed optical fiber infrastructure. While new technologies, such as quantum repeaters, will be needed to realize the long-term goals of quantum networking, IEQNET focuses on leveraging currently available technology (with provision for future upgrades as technology develops). Two of the prime needs in quantum information systems are distributing entanglement and using entanglement to perform quantum state teleportation. These needs should be met while simultaneously allowing higher-power classical signals to share the same fiber, both for the purposes of enabling communications to support quantum applications and for independent coexistence of high data-rate classical channels. In this paper we will introduce our architecture to realize these functions over a metro-scale network, focusing for simplicity on entanglement distribution, and discuss some of the major control and management issues that need to be addressed to enable reliable network operation. While there have been several proof-of-concept demonstrations of deployed quantum communications and networking, over both free space and fiber, in various locations around the globe, see e.g. Refs.~\cite{boaron2018secure,elliott2007darpa,liao2017satellite,peev2009secoqc,pirandola2020advances,sasaki2011field,ursin2007entanglement} for an overview, there is a need to advance entanglement-based technologies to be more integrated into the (classical) networking framework beyond that of previous demonstrations. This includes scaling to more users, reaching longer link distances, allowing coexisting quantum and classical data channels on the same fiber links, as well as optimizing and automating the software control of network operations such as synchronization.

The rest of the paper is organized as follows. In Section~\ref{sec:arch} we describe IEQNET's architecture, design, and implementation. Section~\ref{sec:experiment} provides details on experiment demonstration. Section 4 presents our conclusions and future work.

\section{IEQNET Architecture, Design, and Implementation} \label{sec:arch}

\begin{figure}
    \centering
    \includegraphics[width=\columnwidth]{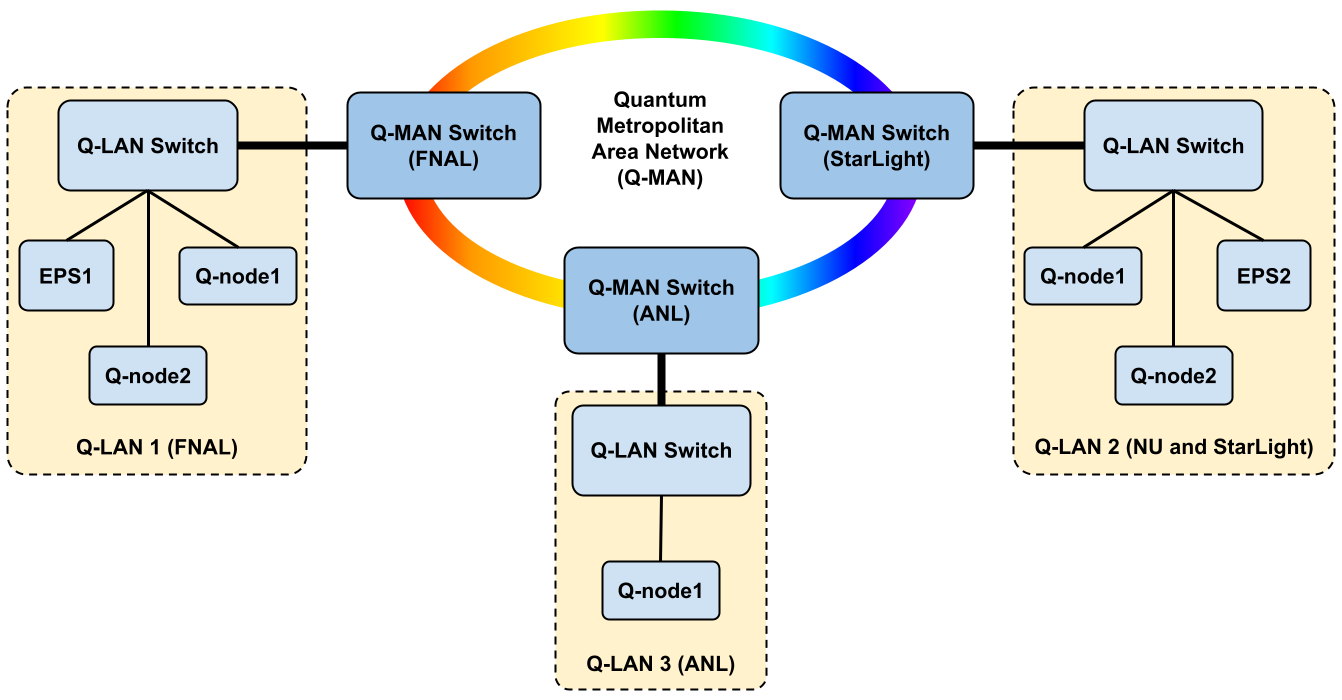}
    \caption{IEQNET topology}
    \label{fig:topo}
\end{figure}

\subsection{IEQNET Topology, Architecture, and Design} \label{sec:topo}
IEQNET consists of multiple sites that are geographically dispersed in the Chicago metropolitan area with sites at Northwestern University (NU), Fermi National Accelerator Laboratory (FNAL), Argonne National Laboratory (ANL), and a Chicago-based data center (StarLight). Each site has one or more quantum nodes (Q-nodes), which can communicate data and generate and/or measure quantum signals. Q-nodes will be connected to software-defined-networking (SDN) enabled optical switches through optical fibers. The optical switches are further connected among one another to form a meshed all-optical network. IEQNET contains three logically independent quantum local area networks (see Figure~\ref{fig:topo}): Q-LAN1 at FNAL, Q-LAN2 at NU and StarLight, and Q-LAN3 at ANL. The Q-LANs are connected by dedicated communication channels and additional dark fibers between FNAL, ANL, and StarLight. In addition, IEQNET has one or multiple shared entangled photon sources (EPS's) and Bell state measurement nodes (BSM nodes), which are also connected to SDN-enabled optical switches through optical fibers. An EPS generates entangled photon pairs at N wavelengths, allowing a maximum of $N/2$ user-pairs to simultaneously share bipartite entangled photons. A BSM node performs \textit{Bell state measurements} and local qubit operations for incoming photon pairs.

Inspired by the Internet architecture, IEQNET implements a similar layered quantum networking architecture, which describes how quantum network functions are vertically composed to provide increasingly complex capabilities. IEQNET's layered quantum networking architecture relies on four key vertical layers. 
\begin{itemize}
    \item \textit{Quantum physical layer} deals with the physical connectivity of two communicating quantum nodes. It defines quantum channel frequencies, signal rates, photon pulses used to represent quantum signals, etc.
    \item \textit{Quantum link layer} is the protocol layer in Q-nodes that handles the transmission of quantum signals and messages across quantum channels. 
    \item \textit{Quantum networking layer} performs wavelength routing and assignment in optical networks to establish quantum paths between quantum nodes. 
    \item \textit{Quantum service layer} provides quantum services, such as entanglement distribution and quantum teleportation to users and applications.
\end{itemize}

Mechanisms for generation, synchronization, and measurement of quantum states in networks are crucial for realizing a multitude of quantum information applications. They can allow high-quality distribution of entanglement throughout the network—provided errors are identified, their magnitude estimated, and steps taken for their correction. Orchestration and control mechanisms are especially important for performing advanced quantum communication tasks, such as quantum teleportation and entanglement swapping, which are based on quantum interference and thus are more susceptible to dynamical processes in a network environment, such as polarization rotations in the fiber channels or electronic control drift from local clock mismatches. IEQNET uses a centralized control approach in which SDN controllers monitor the status of key infrastructure plane metrics (e.g., loss on fiber links, status of optical switches, etc.). IEQNET's control and management software performs such functions as time synchronization, optical path routing and wavelength assignment for quantum and classical channels, channel calibration and optimization, and error detection and feedback.

\begin{itemize}
    \item \textit{Time Synchronization.} Synchronizing remote locations for distribution of entanglement and their use in subsequent applications is crucial for quantum networking. For the fiber channels, this is done by distributing clock pulses in the same fiber as the quantum signals, where permitted. To limit the effect of Raman scattering, which generates spurious photons over a broad range of wavelengths, we set the wavelength of the clock light to be at longer wavelengths compared to that for the quantum signals.
    \item \textit{Routing} is a fundamental network function, and multihop networks require a means of selecting paths through the network. IEQNET's underlying quantum network is a WDM-based all-optical network. We will use SDN technology to perform traditional routing and wavelength assignment (RWA)~\cite{zang2000review} to establish paths between Q-nodes in IEQNET's quantum physical layer. RWA is typically formulated as a multi-commodity ﬂow problem, an NP-hard problem that is typically solved with heuristic algorithms. For IEQNET, we use the following shortest-path RWA (SP-RWA) algorithm as the baseline. We represent the network by an undirected graph $G(V, E)$, where V represents the set of nodes in the graph (Q-nodes, BSM, EPSs, and optical switches) and E represents the set of edges in the graph (optical links). Each edge in the graph (optical link in the network) has the following characteristics that contribute to the computation of the edge’s metric/weight: link length, total number of wavelengths, number of wavelengths available, attenuation, etc. Each node in the graph may also contribute to the edge's metric/weight with attributes such as insertion loss, polarization-dependent loss (PDL), and polarization mode dispersion (PMD).
    \item \textit{Quantum Channel Calibration \& Optimization.} The single-photon nature of quantum communication signals makes them extremely sensitive to noise on the quantum channels. In addition, as mentioned above, protocols such as teleportation require indistinguishability in spectral, temporal, spatial, and polarization properties of the two photons arriving at the BSM node. IEQNET employs several active and automated quantum-channel calibration and optimization mechanisms to minimize quantum-channel loss, reduce background noise, and compensate for polarization and delay drifts. HOM measurement and calibration is used to ensure quantum indistinguishability. The HOM signal provides feedback to compensate for the photons' relative time-of-flight, ensuring stable operation. Active polarization measurement and calibration using coexisting classical signals, such as clock pulses, is used to compensate for polarization drifts in fibers.
\end{itemize}

\subsection{IEQNET Control Plane Implementation} \label{sec:cp-impl}

\begin{figure*}[htb!]
    \centering
    \subfigure[]{\label{fig:cp-impl}\includegraphics[width=.4\textwidth]{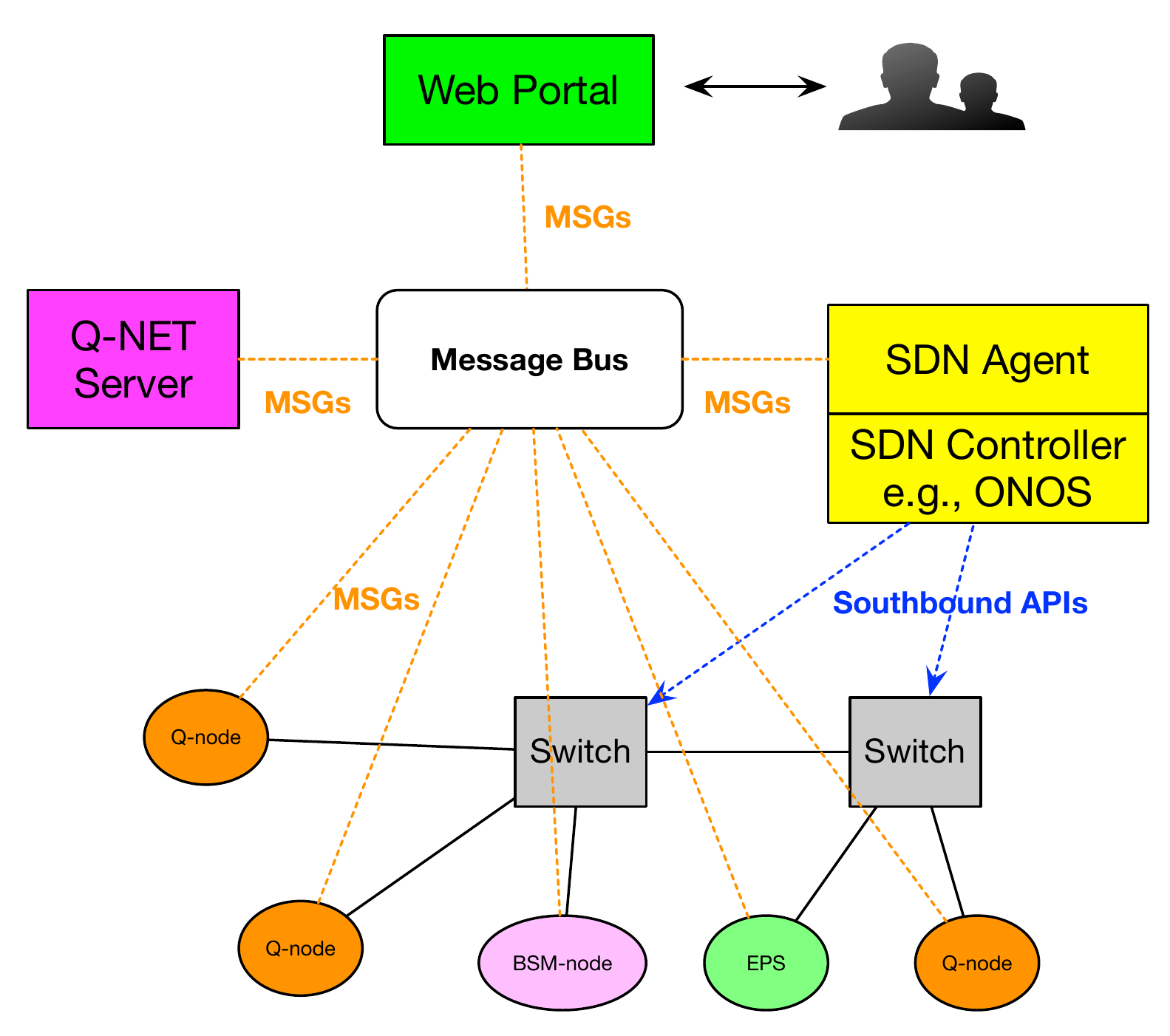}}
    \subfigure[]{\label{fig:qnet-server}\includegraphics[width=.48\textwidth]{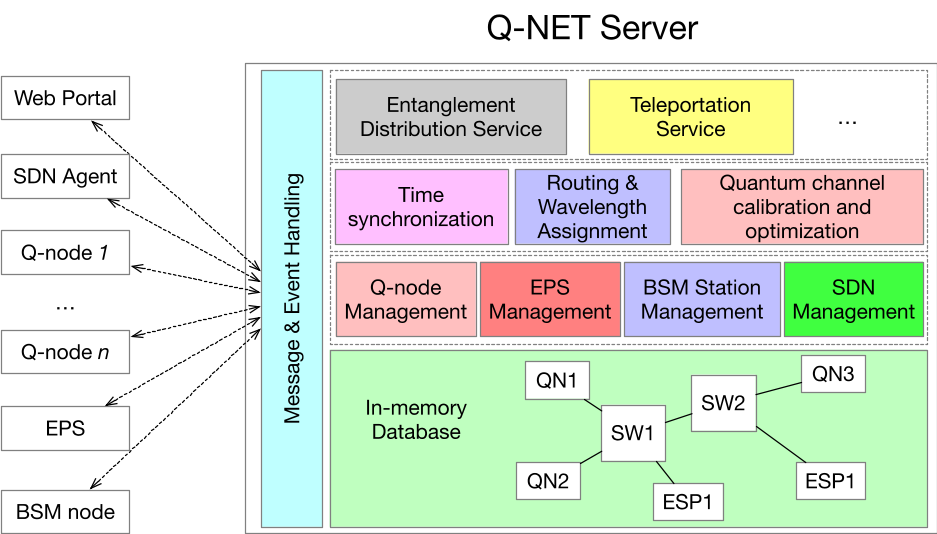}}
	\caption{(a) IEQNET control plane implementation and (b) Q-NET Server block diagram}
	\label{fig:ieqnet-cp}
\end{figure*}

We are working on implementing and deploying an SDN-based, logically centralized control plane for IEQNET. As illustrated in Figure~\ref{fig:ieqnet-cp}, a logically centralized Q-NET server will coordinate all activities in the network. This server will manage and schedule various quantum network resources (Q-nodes, EPS, BSM-nodes, and channels) to perform key control and management functions of quantum networking services, such as entanglement distribution or quantum teleportation.

A web portal will authenticate, authorize, and audit users and applications, and allow them to access IEQNET services. For example, for an entanglement distribution service request, the following information will be conveyed to the Q-NET server via the web portal: the credentials of the task submitter, the Q-nodes involved, and the entanglement distribution requirements, such as qubit type, rate, duration, etc. The Q-NET server will use this information to schedule and broker resources for the task. In addition, users will be able to browse the quantum network topology or monitor the system/site status via the web portal.
An SDN agent will keep track of the quantum topology and traffic status with the aid of SDN controllers. It will also be responsible for reliably updating SDN-enabled switch rules, as requested by the Q-NET server, to assign paths for quantum, clock, and classical signals/messages. SDN controllers are open-source network operating systems, such as ONOS. The SDN agent will access SDN controllers through northbound APIs. One or multiple SDN agents will be deployed, depending on the size of the quantum network.

In this design, the Q-NET server communicates with other entities through a message queuing telemetry transport (MQTT) based message bus. Such a control plane design offers flexibility, robustness, and scalability.

\subsection{IEQNET Control Protocol Suite} \label{sec:protocol}
\begin{figure}
    \centering
    \includegraphics[width=\columnwidth]{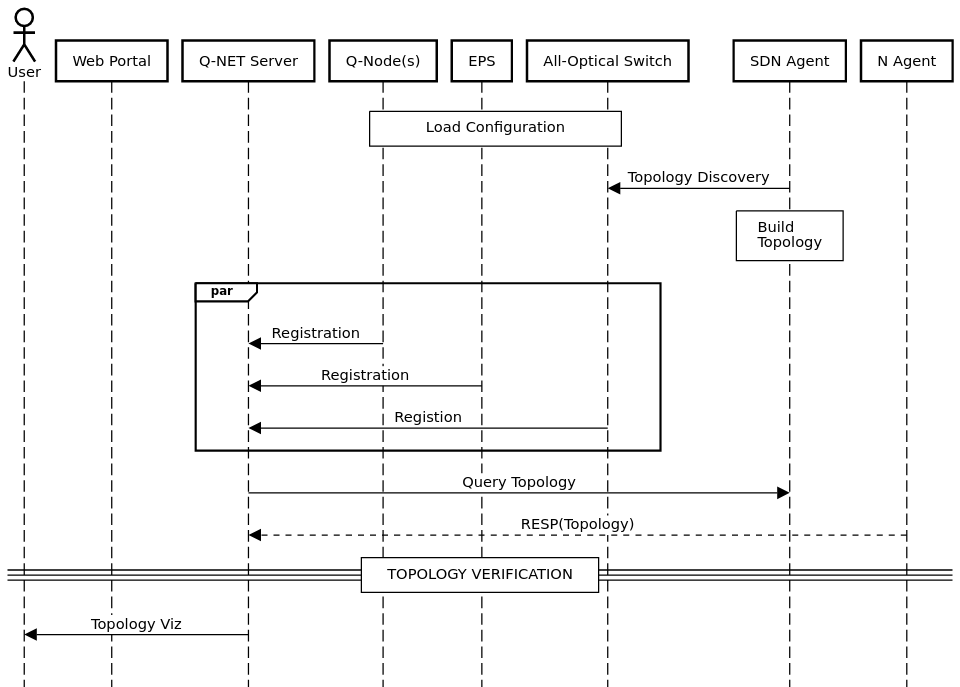}
    \caption{Quantum network resource and discovery protocol}
    \label{fig:discovery}
\end{figure}

\begin{figure}
    \centering
    \includegraphics[width=\columnwidth]{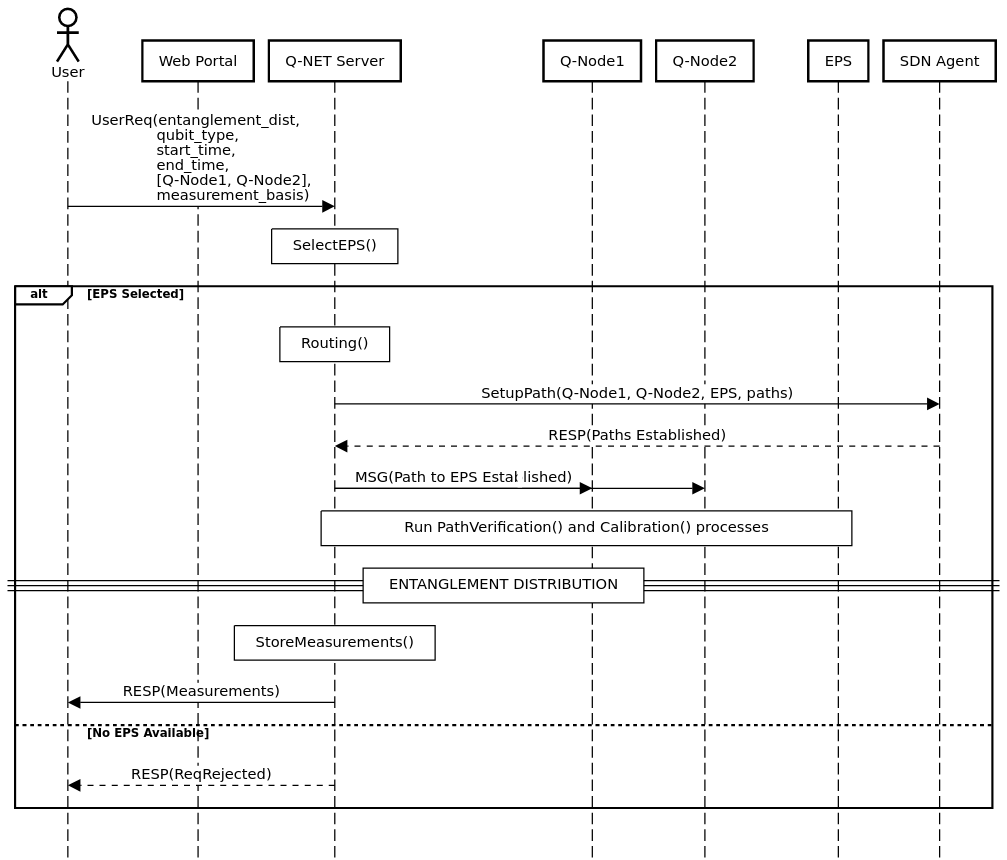}
    \caption{The protocol for handling entanglement distribution requests}
    \label{fig:res-mgmt}
\end{figure}

In this section we describe the IEQNET control protocol suite, which is composed of two main protocols: quantum network resource and topology discovery protocol (see Figure~\ref{fig:discovery}) and the protocol for handling entanglement distribution requests (see Figure~\ref{fig:res-mgmt}). 

Figure~\ref{fig:discovery} shows a sequence diagram of the discovery protocol, which starts by making each quantum networking resource (e.g., Q-Nodes, EPS, BSM-nodes, and switches) load their configuration. The next step is for the SDN agent to discover the network topology through the SDN controller’s southbound API. As all-optical switches are passive devices and thus classical active mechanism for topology discovery will not work, we have extended ONOS’s link discovery service to build the topology from configuration loaded in a tag field on each optical port’s configuration. Quantum network resources register to the Q-NET server by sending their features and connectivity information. The Q-NET server will request the topology from the SDN agent and will subsequently ask the SDN agent to verify the connectivity information provided by individual quantum resources. Once the topology has been verified, the Q-NET server will build a topology graph based on the updated topology and present it to users through the Web portal. This discovery protocol keeps running as quantum network resources can come and go. Furthermore, the SDN agent has the capability to notify the Q-NET server of topology changes asynchronously.

Figure~\ref{fig:res-mgmt} illustrates how IEQNET handles a user entanglement distribution request. The protocol starts with a user requesting entanglement distribution between Q-Node1 and Q-Node2. The Q-NET server will analyze this request and choose an EPS that meets the requirements specified by the user. Upon acceptance of the request, the Q-NET server will execute path routing and wavelength assignment and will establish the paths among involved entities via the SDN agent. Q-NET server will notify Q-Node1 and Q-Node2 when paths are established and initiate path verification, which involves a series of active probes from EPS to Q-Nodes (and vice versa) using both classical and quantum light. After path verification, the Q-NET server will initiate calibration and optimization processes (as described in Section~\ref{sec:topo}) for the requested service. Once all entities send the READY signal to the Q-NET server, the entanglement distribution process starts. Q-Nodes will collect measurements until they have a long enough ebit string for the upper layer application. At that point they will send the END signal to the Q-NET server to stop entanglement distribution. Periodically during entanglement distribution, the Q-NET server will re-initiate the calibration and optimization processes. After the Q-NET server stops the EPS, all measurements will be stored at the Q-NET server and the user will be able to access them through the Web portal.

\section{Experiment Demonstration} \label{sec:experiment}
In this section we present quantum teleportation and coexistence results obtained at experiments performed at both Q-LAN1 and Q-LAN2 of IEQNET. These systems are used to implement and test the IEQNET architecture and processes. We employ time-bin entangled photons to perform teleportation over 44 km of fibers at 1536 nm and achieve above 90\% teleportation fidelities with a semi-autonomous system that can sustain stable operation via modern data-acquisition systems and integrated feedback mechanisms. We also discuss the commissioning results of a clock distribution system, coexisting in the same optical fiber as the quantum channel, and where we measured a time jitter of $\sim$5 ps. These coexisting clock distribution systems will pave the way towards synchronizing multiple nodes in remote locations and thus allowing more complex quantum protocols. Finally, we present preliminary measurements that support network designs with simultaneously coexisting quantum and classical signals on the same fiber connecting the Q-nodes of the network. Our experiments and data analysis are guided and supported by a phenomenological model which can be quickly compared with experimental data.  We discuss the model and its utilization in our results.

\subsection{Quantum Teleportation of Time-bin Qubits and coexistence with telecommunication O-band classical signals at Q-LAN1} \label{sec:teleportation}
Using fiber-coupled devices, including state-of-the-art low-noise superconducting nanowire single photon detectors and off-the-shelf optics, the Q-LAN1 systems achieve quantum teleportation of time-bin qubits at the telecommunication wavelength of 1536.5 nm~\cite{valivarthi2020teleportation}. We measure teleportation fidelities of $>90\%$ that are consistent with an analytical model of our system, which includes realistic imperfections. To demonstrate the compatibility of our setup with deployed quantum networks, we teleport qubits over 22~km of single-mode fiber while transmitting qubits over an additional 22~km of fiber. Our systems, which are compatible with emerging solid-state quantum devices, provide a realistic foundation for a high-fidelity quantum internet with practical devices. This is accomplished using a compact setup of fiber-coupled devices, including low-dark-count single photon detectors and  off-the-shelf optics,  allowing straight-forward reproduction for multi-node networks.

To illustrate network compatibility, teleportation is performed with up to 44 km of single-mode fiber between the qubit generation and the measurement of the teleported qubit, and is facilitated using semi-autonomous control, monitoring, and synchronization systems, with results collected using scalable acquisition hardware. Our system, which operates at a clock rate of 90~MHz, can be run remotely for several days without interruption and yield teleportation rates of a few Hz using the full length of fiber. Our qubits are also compatible with erbium-doped crystals, e.g. Er:Y2SiO5, that is used to develop quantum network devices like memories and transducers~\cite{miyazono2016coupling,welinski2019electron,lauritzen2010telecommunication}. The 1536.5 nm operating wavelength is within the low-loss (C-band) telecommunication window for long-haul communication and where a variety of off-the-shelf equipment is available. We also develop an analytical model of our system, which includes experimental imperfections, predicting that the fidelity can be improved further towards unity by well-understood methods (such as improvement in photon indistinguishability). Our demonstrations provide a step towards a workable quantum network with practical and replicable nodes, such as the ambitious U.S. Department of Energy quantum research network envisioned to link the U.S. National Laboratories.

\begin{figure}
    \centering
    \includegraphics[width=\columnwidth]{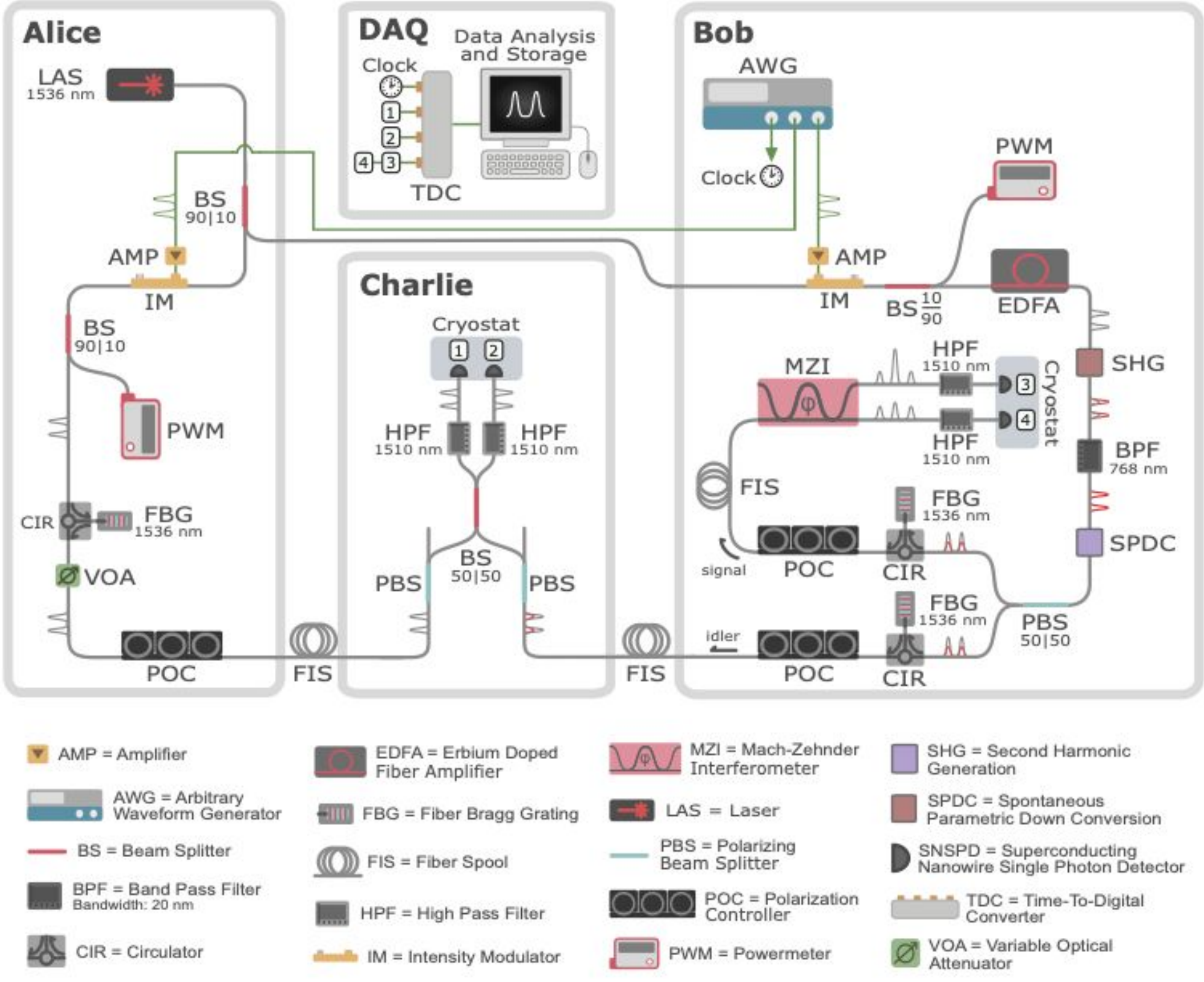}
    \caption{Schematic diagram of the quantum teleportation system consisting of Alice, Bob, Charlie, and the data acquisition (DAQ) subsystems. One cryostat is used to house all SNSPDs, it is drawn as two for ease of explanation. Detection signals generated by each of the SNSPDs are labelled 1-4 and collected at the TDC, with 3 and 4 being time-multiplexed. All individual components are labeled in the legend, with single-mode optical fibers (electronic cables) in grey (green), and with uni- and bi-chromatic (i.e. unfiltered) optical pulses indicated.}
    \label{fig:figure5}
\end{figure}

\begin{figure*}[htb!]
    \centering
    \subfigure[]{\label{fig:figure6a}\includegraphics[width=.48\textwidth, trim=0 0 0 0.75cm, clip]{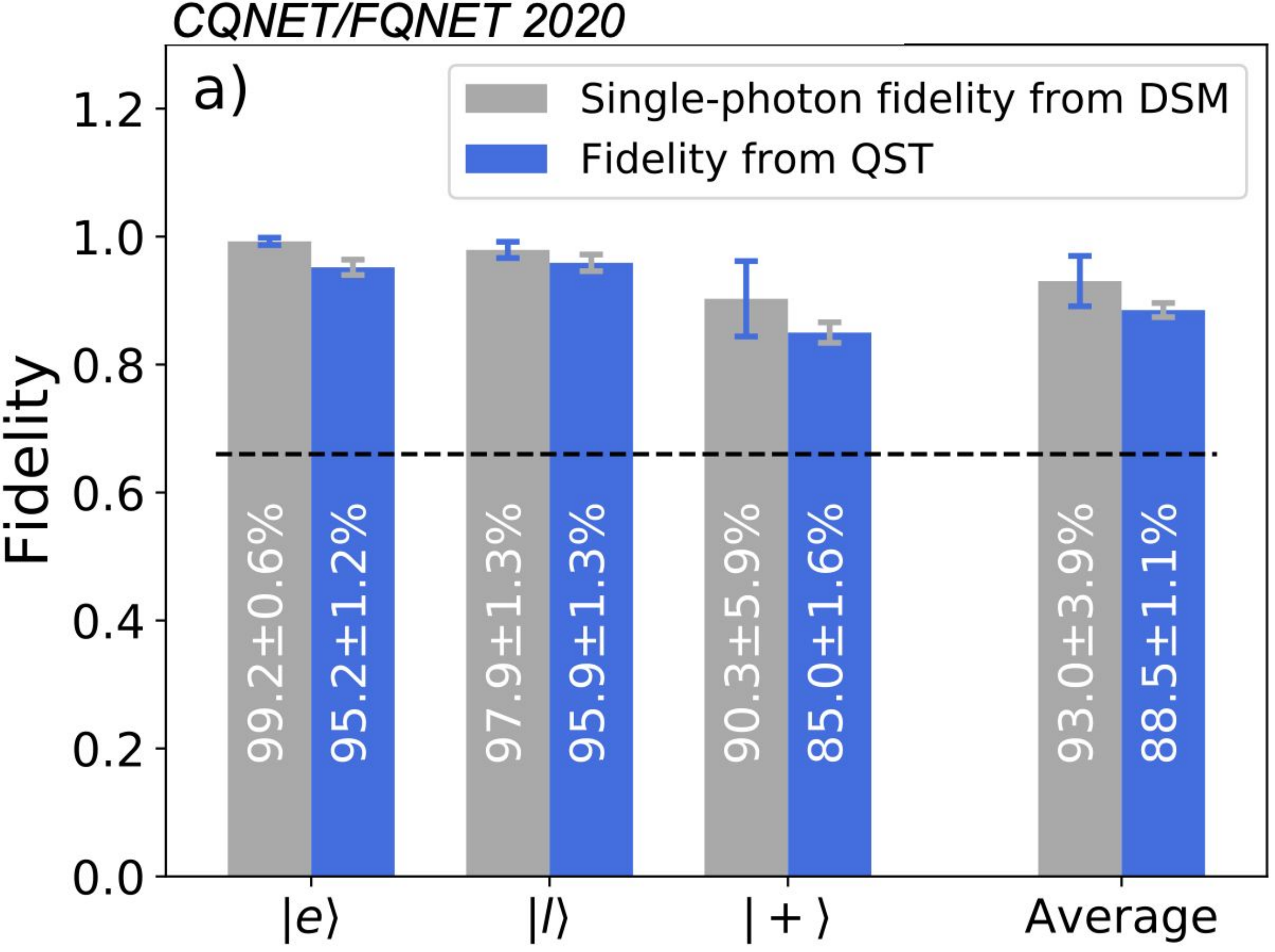}}
    \subfigure[]{\label{fig:figure6b}\includegraphics[width=.46\textwidth]{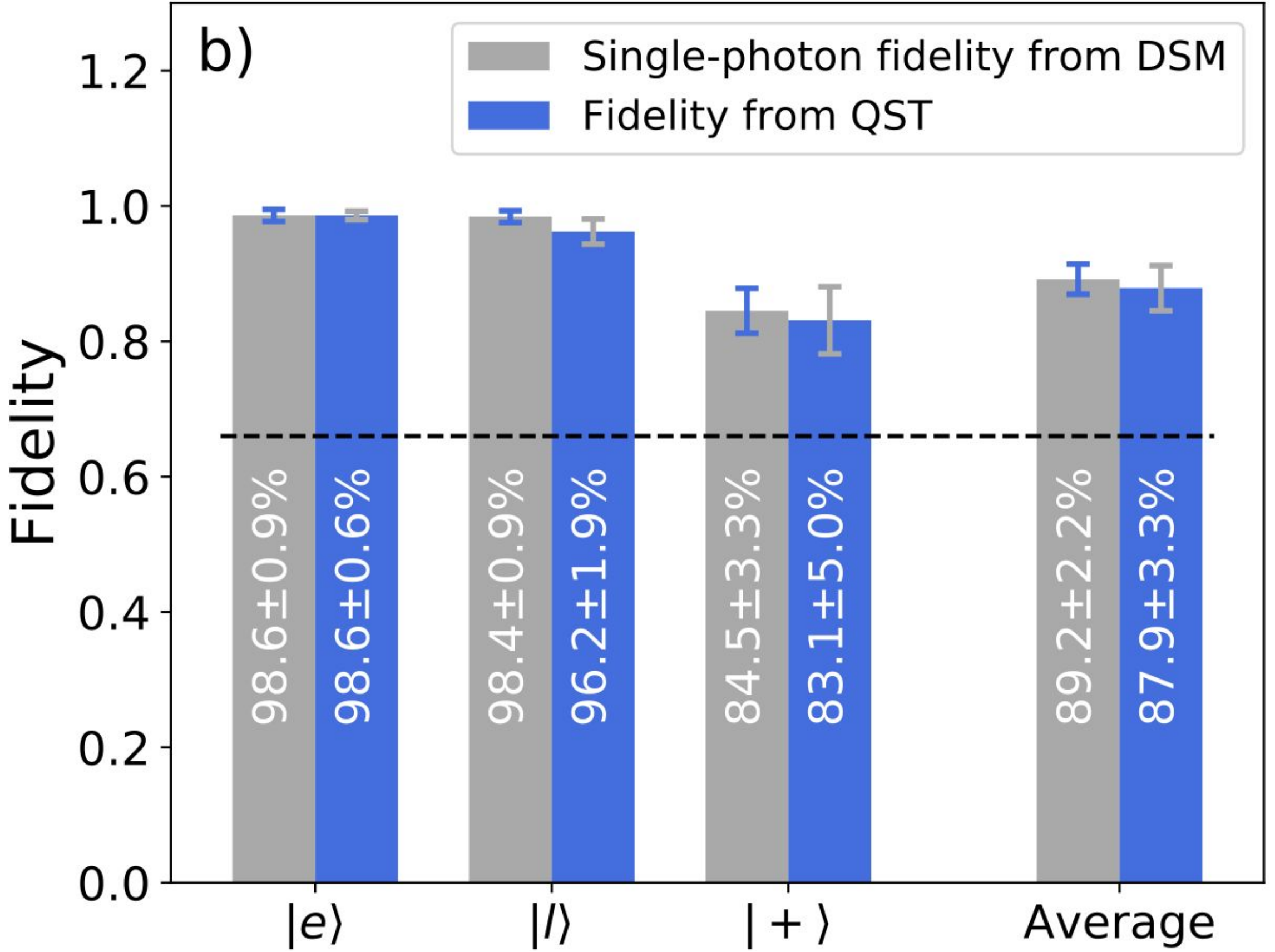}}
	\caption{Quantum teleportation fidelities for $|e>_A$, $|l>_A$, and $|+>_A$, including the average fidelity. The dashed line represents the classical bound. Fidelities using quantum state tomography (QST) are shown using blue bars while the minimum fidelities for qubits prepared using $|n=1>$, $F^d_e$, $F^d_l$, and $F^d_+$, including the associated average fidelity $F^d_avg$, respectively, using a decoy state method (DSM) is shown in grey. Panels a) and b) depict the results without and with additional fiber, respectively. Uncertainties are calculated using Monte-Carlo simulations with Poissonian statistics}
	\label{fig:figure6}
\end{figure*}

The Q-LAN1 fiber-based experimental system, summarized in the diagram of Figure~\ref{fig:figure5}, allows us to demonstrate a quantum teleportation protocol in which a photonic qubit (provided by Alice) is interfered with one member of an entangled photon-pair (from Bob) and projected (by Charlie) onto a Bell-state whereby the state of Alice's qubit can be transferred to the remaining member of Bob's entangled photon pair. Up to 22 (11) km of single mode fiber is introduced between Alice and Charlie (Bob and Charlie), as well as up to another 11~km at Bob, depending on the experiment. All qubits are generated at the clock rate, with all of their measurements collected using a data acquisition (DAQ) system. The measured teleportation fidelities with and without are presented in Figure~\ref{fig:figure6}. Going further, the collaboration is currently working towards time-bin entanglement swapping~\cite{zukowski1993event}. Preliminary results in terms of indistinguishability have been obtained, in the newly commissioned Fermilab FCC Q-node, when interfering the two entanglement photon sources and performing a Hong-Ou-Mandel analysis~\cite{hong1987measurement}.

Another aspect of quantum networking addressed by the Q-LAN1 experiments is clock distribution. Since photons are identified by recording their times of generation and detection, each with respect to a local (node-based) clock, such variations can lead to misidentification of photons. To avoid this, the variations can be accounted for by adjusting the phases of each local clock using strong optical pulses that are co-transmitted with individual photons. We developed such a clock distribution system, in which O-band light is directed to two independent nodes to adjust their local clocks while C-band photon pairs, originating from a source based on spontaneous parametric down-conversion, are co-transmitted. We find that the clock distribution system can allow for high-fidelity qubit distribution despite the presence of (Raman) noise. Furthermore, we observe little additional timing jitter between clocks at the central and end nodes, suggesting our method can be used for high-rate networks. The deployment of such a system will pave the road towards implementing important multi-node quantum functions towards scalable networks. 

\subsection{Coexistence Studies of Quantum-Entangled and Classical Light on Real-World Installed Fiber at Q-LAN2} \label{sec:coexistence}
Future quantum networks will leverage the traditional fiber optical infrastructure as much as possible to gain the cost efficiency in deploying such networks. This means that quantum signals will likely coexist with classical communications that carry either information related (e.g., synchronization signals or quantum-protocol related data) or unrelated (e.g., independent high-rate data communication) to the quantum channel. Anticipating this need, there have been many experimental studies of quantum-classical coexistence in standard telecom fibers using attenuated laser light as proxy for quantum signals. Most such studies have been in the context of quantum key distribution (QKD), and rightfully so, because QKD is the most advanced application of quantum communications that is closest to real-world deployment~\cite{mao2018integrating}. However, quantum network functions beyond QKD, such as teleportation, quantum repeaters, etc., will require the distribution of quantum entanglement in real-world scenarios. Entangled light is not only required for such quantum communication functions, but it also has an inherent robustness to background noise due to the use of coincidence detection. In fibers carrying multiplexed quantum and classical signals, the dominant background noise is expected to be Raman-scattering from the copropagating classical light channels. One approach to suppressing such Raman noise is to place the quantum signals in the 1310-nm telecom O-band and the classical signals far away in the traditional 1550-nm telecom C-band~\cite{chapuran2009optical}, which ensures that the sub-photon-level quantum signals are far detuned ($\sim$35 THz) on the anti-Stokes side of the much stronger classical light. In addition, by tightly filtering the quantum signals in time and frequency domains, we have shown in recent experiments (described below) that polarization-entangled quantum light can co-propagate with milliwatt-level classical light over 45~km of installed underground fiber that presents about 20~dB of loss to the quantum signals. Such classical light levels are consistent with modern high-rate communication channels, and far exceed another recent entangled and classical light coexistence experiment that used C-band for both the quantum and classical channels~\cite{yuan2019quantum}.

\begin{figure*}
    \centering
    \includegraphics[width=\textwidth]{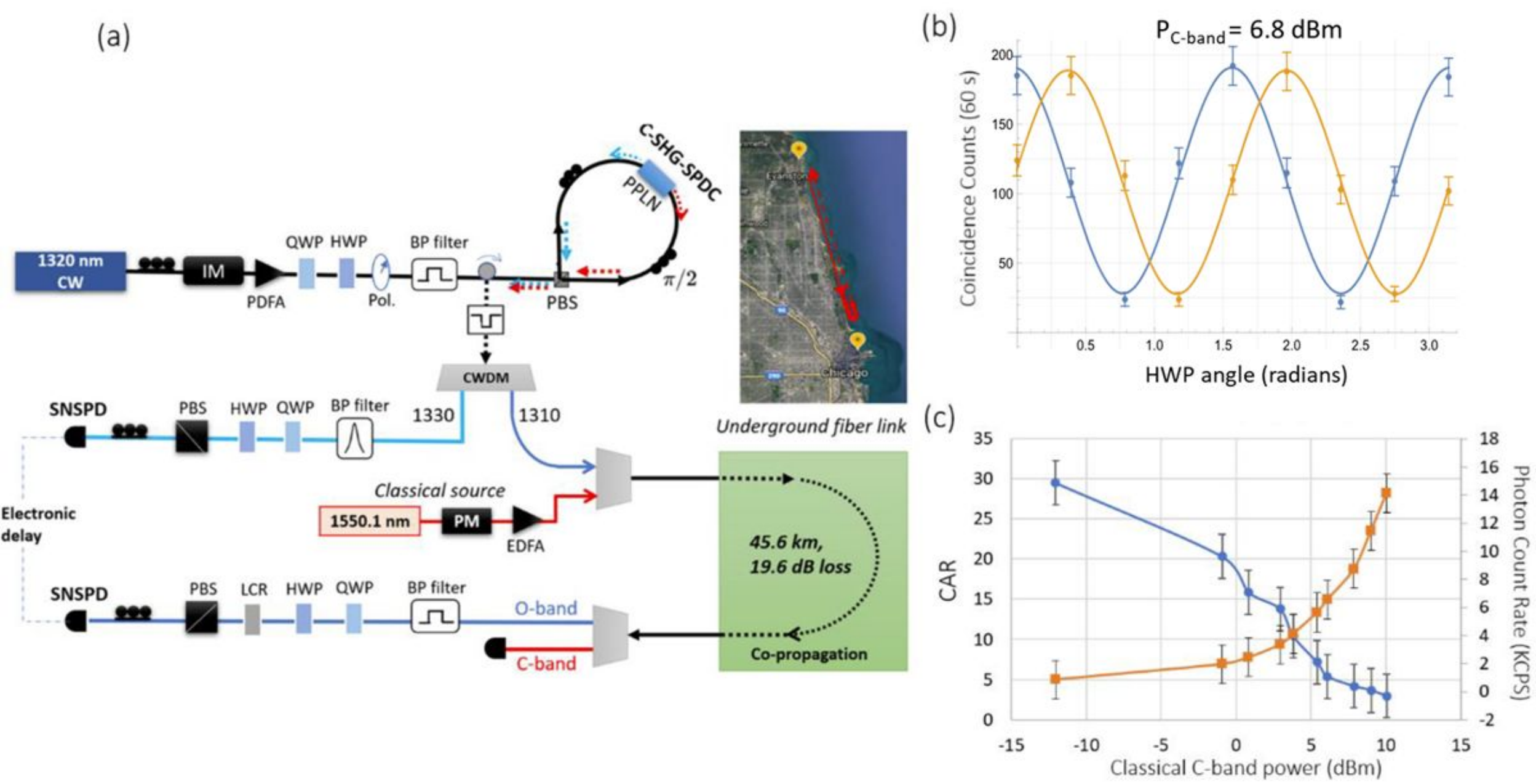}
    \caption{(a) Experimental diagram. (b) Coincidence counts as a function of relative HWP angle between signal and idler polarization analyzers for a co-propagating classical power of $P_{C-band} = 6.8~dBm$. Data in blue is for the horizontal-vertical (HV) basis and that in yellow is for the diagonal-antidiagonal (DA) basis. (c) CAR (blue) and single photon count rates (orange) as a function of C-band power (KCPS = kilo-counts per second)}
    \label{fig:figure7}
\end{figure*}

In our experiments, the use of O-band/C-band quantum/classical wavelength allocation allows us to achieve much higher co-propagating classical powers (about 7 dBm) while still maintaining high visibility ($>$71\%) in polarization entanglement two-photon interference. Figure~\ref{fig:figure7}(a) shows a schematic of our experiment. We send one photon of the polarization-entangled photon pair over a 45.6~km loop of underground installed fiber. The underground fiber link connects the Quantum Communications Laboratory at Northwestern University in Evanston to the Starlight Communications Facility located on the Northwestern Campus in Chicago (link distance of 22.8~km) where we loop the co-propagating light back to the Evanston laboratory for characterization with polarization analyzers and low-dark-count superconducting nanowire single photon detectors (SNSPDs). The measured loss in the underground fiber link is 0.43~dB/km at 1310~nm, which is higher than expected in modern fibers typically used in laboratory experiments, making the equivalent loss closer to a 60 km distance if newer fibers were used. 

We generate the quantum signal via cascaded second harmonic generation-spontaneous parametric down conversion (c-SHG-SPDC) in a single periodically-poled lithium-niobate waveguide (PPLN)~\cite{arahira2011generation}. The waveguide is phase matched for SHG of the 1320 nm pump pulse train at 417 MHz repetition rate with 80 ps pulse-width. We place the waveguide inside a polarization Sagnac loop to generate polarization entangled photon pairs. The pump light entering the loop is split by a polarizing beam splitter (PBS) into two counter-propagating directions. The c-SHG-SPDC generates broad-bandwidth quantum amplitudes for photon pairs centered around 1320 nm in both directions, and upon recombination at the PBS the two-photon quantum state becomes polarization-entangled. We separate the signal/idler photons into bands using a standard coarse wavelength division multiplexer (CWDM). The CWDM outputs 20 nm wide bands with center wavelengths of 1310 and 1330 nm. We send the 1310 nm channel over the 45.6 km underground fiber link while keeping the 1330 nm channel locally. For the classical channel, we amplify the C-band light from a phase modulated (PM) CW laser at 1550.1 nm with an erbium-doped fiber amplifier (EDFA) and multiplex it into the underground fiber to co-propagate with the O-band quantum signal. We phase modulate the C-band light to broaden its spectrum in order to emulate a data channel and inhibit stimulated Brillouin scattering. At the receiver, we demultiplex out the C-band return light and filter the signal and idler with 100 GHz bandpass (BP) filters at 1306.5 and 1333.5 nm, respectively. 

We detect both the signal and idler photons by SNSPDs, which are followed by a time-tagging correlation detection system. We apply an electronic delay between the two channels to account for the fiber delay and perform coincidence measurements. We use a pair of polarization analyzers consisting of a quarter-wave plate (QWP), a half-wave plate (HWP), and a PBS to make arbitrary projections of the incoming entangled-photon polarizations. Figure~\ref{fig:figure7}(b) shows two-photon interference fringes after the transmitted photon has propagated alongside 6.8 dBm of C-band launch power, where a visibility of 77\% is observed in the HV basis and 74\% in the DA basis. Both values are $>$71\% and thus fall in the nonclassical regime of two-photon interference. Figure~\ref{fig:figure7}(c) shows a plot of the coincidence-to-accidental ratio (CAR) as a function of the co-propagating C-band power for an entangled pair rate higher than in Figure~\ref{fig:figure7}(b). CAR quantifies the signal-to-noise ratio of the coincidence detection of photon pairs, where CAR $>$6 is desired for most applications. CAR falls as the co-propagating power is increased, as expected. Given the $\sim$0.5 ns temporal correlation window, frequency filtering to $<$5 GHz could presumably reduce Raman noise by another factor of $\sim$20, which would allow for even higher co-propagating powers ($>$10 dBm) to be used.

The above results clearly demonstrate that O-band/C-band quantum/classical wavelength allocation along with temporal and spectral filtering are useful noise mitigation methods for coexistence scenarios in fiber-optic quantum networking. Since future quantum networks will require entanglement distribution, we expect that the noise mitigation methods presented here will be useful when integrating quantum network functionality into the installed fiber optical infrastructure.
\section{Conclusion} \label{sec:conclusion}
We have developed a network architecture which aids in developing free-running,  practical, quantum network controls and demonstrating metro-scale quantum operations including entanglement distribution and eventually quantum teleportation.
We have performed experiments on quantum teleportation and coexistence of quantum and classical information on deployed fiber.
Our complementary approaches of distributing quantum and classical signals in the telecommunication C- and O-band, or vice versa, allow greater network flexibility and testing under varied conditions and use cases. For example, quantum communication in the C-band is suitable for long-haul links, while the utility of qubits in the O-band are not restricted by noise from strong classical light. Importantly, these two approaches are not mutually exclusive but can be linked using quantum teleportation. This will ensure quantum information can be routed between IEQNET Q-LANs, opening the door for real-world testing of hybrid quantum physical layer architectures.

\section*{Acknowledgment}
IEQNET is funded by the Department of Energy's Advanced Scientific 
Computing Research Transparent Optical Quantum Networks for 
Distributed Science program, but no government endorsement is implied. 
Fermilab is managed by Fermi Research Alliance, LLC under 
Contract No. DE-AC02-07CH11359 with the U.S. Department of Energy, 
Office of Science, Office of High Energy Physics.

\bibliographystyle{IEEEtran}
\bibliography{bibliography.bib}

\begin{thebibliography}{10}
\providecommand{\url}[1]{#1}
\csname url@samestyle\endcsname
\providecommand{\newblock}{\relax}
\providecommand{\bibinfo}[2]{#2}
\providecommand{\BIBentrySTDinterwordspacing}{\spaceskip=0pt\relax}
\providecommand{\BIBentryALTinterwordstretchfactor}{4}
\providecommand{\BIBentryALTinterwordspacing}{\spaceskip=\fontdimen2\font plus
\BIBentryALTinterwordstretchfactor\fontdimen3\font minus
  \fontdimen4\font\relax}
\providecommand{\BIBforeignlanguage}[2]{{%
\expandafter\ifx\csname l@#1\endcsname\relax
\typeout{** WARNING: IEEEtran.bst: No hyphenation pattern has been}%
\typeout{** loaded for the language `#1'. Using the pattern for}%
\typeout{** the default language instead.}%
\else
\language=\csname l@#1\endcsname
\fi
#2}}
\providecommand{\BIBdecl}{\relax}
\BIBdecl

\bibitem{boaron2018secure}
A.~Boaron, G.~Boso, D.~Rusca, C.~Vulliez, C.~Autebert, M.~Caloz, M.~Perrenoud,
  G.~Gras, F.~Bussi{\`e}res, M.-J. Li \emph{et~al.}, ``Secure quantum key
  distribution over 421 km of optical fiber,'' \emph{{Physical Review
  Letters}}, vol. 121, no.~19, p. 190502, 2018.

\bibitem{elliott2007darpa}
C.~Elliott and H.~Yeh, ``Darpa quantum network testbed,'' BBN TECHNOLOGIES
  CAMBRIDGE MA, Tech. Rep., 2007.

\bibitem{liao2017satellite}
S.-K. Liao, W.-Q. Cai, W.-Y. Liu, L.~Zhang, Y.~Li, J.-G. Ren, J.~Yin, Q.~Shen,
  Y.~Cao, Z.-P. Li \emph{et~al.}, ``Satellite-to-ground quantum key
  distribution,'' \emph{Nature}, vol. 549, no. 7670, pp. 43--47, 2017.

\bibitem{peev2009secoqc}
M.~Peev, C.~Pacher, R.~All{\'e}aume, C.~Barreiro, J.~Bouda, W.~Boxleitner,
  T.~Debuisschert, E.~Diamanti, M.~Dianati, J.~Dynes \emph{et~al.}, ``The
  secoqc quantum key distribution network in vienna,'' \emph{New Journal of
  Physics}, vol.~11, no.~7, p. 075001, 2009.

\bibitem{pirandola2020advances}
S.~Pirandola, U.~L. Andersen, L.~Banchi, M.~Berta, D.~Bunandar, R.~Colbeck,
  D.~Englund, T.~Gehring, C.~Lupo, C.~Ottaviani \emph{et~al.}, ``Advances in
  quantum cryptography,'' \emph{Advances in Optics and Photonics}, vol.~12,
  no.~4, pp. 1012--1236, 2020.

\bibitem{sasaki2011field}
M.~Sasaki, M.~Fujiwara, H.~Ishizuka, W.~Klaus, K.~Wakui, M.~Takeoka, S.~Miki,
  T.~Yamashita, Z.~Wang, A.~Tanaka \emph{et~al.}, ``Field test of quantum key
  distribution in the tokyo qkd network,'' \emph{{Optics Express}}, vol.~19,
  no.~11, pp. 10\,387--10\,409, 2011.

\bibitem{ursin2007entanglement}
R.~Ursin, F.~Tiefenbacher, T.~Schmitt-Manderbach, H.~Weier, T.~Scheidl,
  M.~Lindenthal, B.~Blauensteiner, T.~Jennewein, J.~Perdigues, P.~Trojek
  \emph{et~al.}, ``Entanglement-based quantum communication over 144 km,''
  \emph{{Nature Physics}}, vol.~3, no.~7, pp. 481--486, 2007.

\bibitem{zang2000review}
H.~Zang, J.~P. Jue, B.~Mukherjee \emph{et~al.}, ``A review of routing and
  wavelength assignment approaches for wavelength-routed optical wdm
  networks,'' \emph{{Optical Networks Magazine}}, vol.~1, no.~1, pp. 47--60,
  2000.

\bibitem{valivarthi2020teleportation}
R.~Valivarthi, S.~I. Davis, C.~Pe{\~n}a, S.~Xie, N.~Lauk, L.~Narv{\'a}ez, J.~P.
  Allmaras, A.~D. Beyer, Y.~Gim, M.~Hussein \emph{et~al.}, ``Teleportation
  systems toward a quantum internet,'' \emph{PRX Quantum}, vol.~1, no.~2, p.
  020317, 2020.

\bibitem{miyazono2016coupling}
E.~Miyazono, T.~Zhong, I.~Craiciu, J.~M. Kindem, and A.~Faraon, ``Coupling of
  erbium dopants to yttrium orthosilicate photonic crystal cavities for on-chip
  optical quantum memories,'' \emph{Applied Physics Letters}, vol. 108, no.~1,
  p. 011111, 2016.

\bibitem{welinski2019electron}
S.~Welinski, P.~J. Woodburn, N.~Lauk, R.~L. Cone, C.~Simon, P.~Goldner, and
  C.~W. Thiel, ``Electron spin coherence in optically excited states of
  rare-earth ions for microwave to optical quantum transducers,''
  \emph{{Physical Review Letters}}, vol. 122, no.~24, p. 247401, 2019.

\bibitem{lauritzen2010telecommunication}
B.~Lauritzen, J.~Min{\'a}{\v{r}}, H.~De~Riedmatten, M.~Afzelius, N.~Sangouard,
  C.~Simon, and N.~Gisin, ``Telecommunication-wavelength solid-state memory at
  the single photon level,'' \emph{{Physical Review Letters}}, vol. 104, no.~8,
  p. 080502, 2010.

\bibitem{zukowski1993event}
M.~Zukowski, A.~Zeilinger, M.~A. Horne, and A.~K. Ekert, ``"
  event-ready-detectors" bell experiment via entanglement swapping.''
  \emph{{Physical Review Letters}}, vol.~71, no.~26, 1993.

\bibitem{hong1987measurement}
C.-K. Hong, Z.-Y. Ou, and L.~Mandel, ``Measurement of subpicosecond time
  intervals between two photons by interference,'' \emph{{Physical Review
  Letters}}, vol.~59, no.~18, p. 2044, 1987.

\bibitem{mao2018integrating}
Y.~Mao, B.-X. Wang, C.~Zhao, G.~Wang, R.~Wang, H.~Wang, F.~Zhou, J.~Nie,
  Q.~Chen, Y.~Zhao \emph{et~al.}, ``Integrating quantum key distribution with
  classical communications in backbone fiber network,'' \emph{{Optics
  Express}}, vol.~26, no.~5, pp. 6010--6020, 2018.

\bibitem{chapuran2009optical}
T.~Chapuran, P.~Toliver, N.~Peters, J.~Jackel, M.~Goodman, R.~Runser,
  S.~McNown, N.~Dallmann, R.~Hughes, K.~McCabe \emph{et~al.}, ``Optical
  networking for quantum key distribution and quantum communications,''
  \emph{New Journal of Physics}, vol.~11, no.~10, p. 105001, 2009.

\bibitem{yuan2019quantum}
C.~Yuan, H.~Yu, Z.~Zhang, Y.~Wang, H.~Li, L.~You, Y.~Wang, H.~Song, G.~Deng,
  and Q.~Zhou, ``Quantum entanglement distribution coexisting with classical
  fiber communication,'' in \emph{Asia Communications and Photonics
  Conference}.\hskip 1em plus 0.5em minus 0.4em\relax Optical Society of
  America, 2019, pp. T2F--2.

\bibitem{arahira2011generation}
S.~Arahira, N.~Namekata, T.~Kishimoto, H.~Yaegashi, and S.~Inoue, ``Generation
  of polarization entangled photon pairs at telecommunication wavelength using
  cascaded $\chi$ (2) processes in a periodically poled {LiNbO3} ridge
  waveguide,'' \emph{{Optics Express}}, vol.~19, no.~17, pp. 16\,032--16\,043,
  2011.

\end{thebibliography}

\end{document}